# *Low Temperature Spin Seebeck Effect in Non-Magnetic Vanadium Dioxide*

Renjie Luo[1], Tanner J. Legvold[1], Liyang Chen[2], Henry Navarro[3], Ali C. Basaran[3], Deshun Hong[4], Changjiang Liu[4], Anand Bhattacharya[4], Ivan K. Schuller[3], Douglas Natelson[1,5]

[1]*Department of Physics and Astronomy, Rice University, Houston TX 77005, USA*

[2]*Applied Physics Graduate Program, Rice Quantum Institute, Rice University, Houston TX 77005, USA*

[3]*Department of Physics and Center for Advanced Nanoscience, University of California-San Diego, La Jolla, CA 92093, USA*

[4]*Materials Science Division, Argonne National Laboratory, Lemont, IL 60439, USA*

[5]*Department of Electrical and Computer Engineering and Department of Materials Science and NanoEngineering, Rice University, Houston TX 77005, USA*

**Abstract**

The spin Seebeck effect (SSE) is sensitive to thermally driven magnetic excitations in magnetic insulators. Vanadium dioxide in its insulating low temperature phase is expected to lack magnetic degrees of freedom, as vanadium atoms are thought to form singlets upon dimerization of the vanadium chains. Instead, we find a paramagnetic SSE response in $VO_2$ films that grows as the temperature decreases below 50 K. The field and temperature dependent SSE voltage is qualitatively consistent with a general model of paramagnetic SSE response and inconsistent with triplet spin transport. Quantitative estimates find a spin Seebeck coefficient comparable in magnitude to that observed in strongly magnetic materials. The microscopic nature of the magnetic excitations in $VO_2$ requires further examination.

**I. INTRODUCTION**

The spin Seebeck effect (SSE) [1–9] uses a temperature gradient to generate a net current of mobile spin-carrying excitations in a magnetically active material and has proven useful in characterizing angular momentum transport in magnetic insulators [4]. The SSE has been extensively studied in various magnetic materials, including ferrimagnets [10–13], ferromagnets [14,15], and antiferromagnets [16–19], where magnon excitations and their transport [6–8] are believed to play the essential role. The SSE has also revealed spin transport via paramagnons and other more exotic mobile excitations in paramagnets known to contain interacting local magnetic moments. The paramagnetic SSE was first observed in $Gd_3Ga_5O_{12}$ (gadolinium gallium garnet, GGG, a geometrically frustrated magnetic material) and $DyScO_3$ (at temperatures above its Néel temperature of 3.1 K) [20], where conventional magnon theory fails. In GGG, short-range order and field-induced long-range correlations [20,21] are thought to contribute to the SSE, despite the lack of long-range order. Later, paramagnetic SSE was observed in the paramagnetic phase of ferromagnets above $T_C$ (*e.g.*, $CoCr_2O_4$ [22], and $CrSiTe_3$ and $CrGeTe_3$ [14]) and antiferromagnets above $T_N$ (*e.g.*, $DyScO_3$ [20], $FeF_2$ [23], $RbMnF_3$ [24]). The SSE from paramagnets was also found in the one-dimensional (1D) quantum spin liquid (QSL) system $Sr_2CuO_3$ [25,26] and the spin-Peierls system $CuGeO_3$ [27], associated with the thermal generation of more exotic spin excitations, such as spinons in the 1D QSL and mobile triplets (triplons) in the spin-Peierls system, respectively. Additionally, the spin-gapped quantum magnet, $Pb_2V_3O_9$ [28], showed SSE at low temperatures, with a peak behavior near the critical field for the Bose–Einstein condensation of triplons. In all of these paramagnetic insulators that exhibit SSE response, local moments are present and coupled by strong exchange interactions.

Recently, a general theoretical model of the paramagnetic SSE was developed based on the temperature difference between spins in the insulating paramagnet and the conduction electrons in the spin-orbit metal [29]. While not accounting for bulk SSE in the paramagnet, this model qualitatively reproduces the field-induced reduction of the SSE observed at high fields and low temperatures in the Pt/GGG system.

Strong electronic correlations can lead to the emergence of local moments and unusual spin excitations. Vanadium dioxide ($VO_2$) is a paradigmatic example of a correlated transition metal oxide, with a metal-insulator phase transition at ~345 K in bulk, between a high-temperature rutile metallic phase and a low-temperature monoclinic insulating phase [30–32]. Thermodynamic arguments [33], quantum Monte Carlo calculations [34] and low-frequency Raman spectra [35] indicate that, in the monoclinic phase, the vanadium ions form dimers, each of which comprises a spin singlet in the ground state (as shown in Fig. 1b). As a result, insulating VO$_2$ is expected to be nonmagnetic in the sense of lacking local moments. In practice, VO$_2$ is paramagnetic throughout the range of temperatures covering the metallic and insulating states [36]. The deviation from Curie law susceptibility at low temperatures (see Fig. 5) has been suggested to result from paramagnetic contributions from unpaired electrons created by thermal excitation of triplet states [37]. Previous studies of nonlocal SSE in VO$_2$ [38] showed that at low temperatures, the thermally generated excitations could transport angular momentum.

In this work, we measure a readily detectable longitudinal spin Seebeck response in the nonmagnetic insulating phase of VO$_2$ films at low temperatures. The longitudinal spin Seebeck effect (LSSE) voltage grows linearly with increasing field at low fields but experiences a field-induced reduction at high fields and the lowest temperatures, qualitatively consistent with the recent model of paramagnetic SSE response [29]. The LSSE shows the expected angular dependence with the in-plane field orientation and is linear in the heater power. When the heater power is held constant, the magnitude of LSSE voltage peaks with increasing temperature. The sign of the LSSE response is not consistent with that expected for a triplon-dominated SSE, in which mobile triplet excitations are the angular momentum carriers. The magnitude of SSE in VO$_2$ is comparable to that in Y$_3$Fe$_5$O$_{12}$ (YIG) [39], a paradigmatic ferrimagnetic insulator that exhibits magnon-mediated SSE. The magnetic degrees of freedom in the VO$_2$ and the mechanism behind such an unexpectedly large paramagnetic SSE call for further studies.

**II. EXPERIMENAL SETUP AND METHODS**

In the on-chip-heating geometry of the LSSE [40], a current flowing through a heater wire is driven at angular frequency $\omega$, creating a temperature gradient normal to the sample surface with an AC component at $2\omega$. This drives an angular momentum current, and a voltage at $2\omega$ can be detected at a nearby inverse spin Hall (ISH) detector made from a strong spin-orbit metal (*e.g.*, Pt) for a properly oriented magnetization of the insulator. Single-phase epitaxially grown VO$_2$ thin films with different thicknesses (50 nm, 100 nm, 250 nm, 400 nm) were deposited on 7×12 mm$^2$ [1$\bar{1}$02] r-plane Al$_2$O$_3$ substrates using RF magnetron sputtering from a V$_2$O$_3$ target (99.9% purity) at a substrate temperature of 520 ℃ in Ar/O$_2$ mixture (8 % O$_2$) at 3.7 mTorr [41]. The substrate was later cooled down to 20 °C at a rate of 12 °C min$^{-1}$. X-ray diffraction measurements confirmed single-phase, textured growth along (100) for VO$_2$. A schematic of the device is presented in Fig. 1a and a photography is shown in Fig. 13.

Photolithography, magnetron sputtering, and liftoff were used to prepare the Pt (W) wire (800 μm long, 10 μm wide, 10 nm thick) on the VO$_2$ film surface. A lithographically defined SiO$_x$ layer with a thickness of 100 nm and a Au heater wire (1300 μm long, 10 μm wide, 50 nm thick) were fabricated on the top of the Pt (W) wire by e-beam deposition and liftoff. The SiO$_x$ layer electrically isolates the Au heater and the Pt (W) wire. An AC current at angular frequency $\omega = 2\pi \times (7.7\ \text{Hz})$ is driven through the Au wire, while the voltage across the Pt (W) wire is measured at $2\omega$ using a lock-in amplifier. The measurements are performed as a function of temperature and field in a Quantum Design Physical Property Measurement System (9T-PPMS) and 14T-DynaCool equipped with a rotation stage.

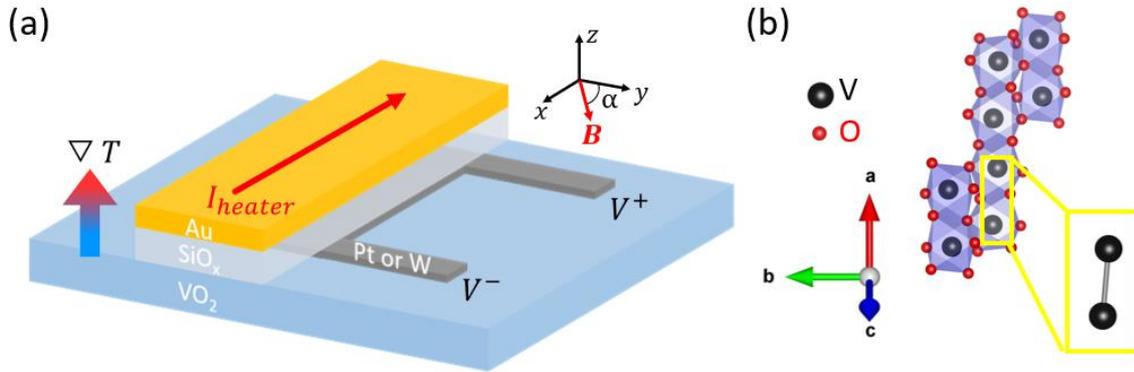

**Fig. 1**. (a) Schematic of local spin Seebeck measurement. An AC heater current produces an oscillating $z$-directed temperature gradient. A vertical ($z$-directed) thermal spin current with the $y$ component of the paramagnetic magnetization of the VO$_2$ could produce an ISH voltage along the $x$-directed strong spin-orbit metal wire. (b) Crystal structure of VO$_2$ in the low temperature, insulating monoclinic phase. The inset shows parallel zigzag chains each consisting of V-V dimers in this phase. Films in this work have the V chains oriented along the $z$ direction, parallel to the applied temperature gradient. The crystal structure is generated by VESTA [42].

## III. RESULTS

The magnetic field dependence of the second harmonic signals is shown for Pt/VO$_2$ (100 nm thick) (Fig. 2a, 2b) and W/VO$_2$ (100 nm thick) (Fig. 2c) for in-plane field oriented at $\alpha = 0°$ for different selected temperatures, with direction and polarity defined as in Fig. 1. At $T = 50$ K, we observed almost no voltage signal (Fig. 2a). With decreasing $T$, a clear $V_{2\omega}$ signal appears, with a sign that changes with respect to the $B$ direction, reflecting the symmetry of the ISHE. When the temperature is above 5 K, the signal magnitude increases monotonically with increasing field, linearly near $B = 0$ T, resembling the *M(H)* curve (Fig. 5a); whereas below 5 K, the signal takes the maximum value at a certain field (for example, 5.2 T at 2.5 K) (Fig. 2b). The voltage responses for devices with Pt and W detectors are of opposite signs (shown in Figs. 2a,c,f), as expected for a genuine spin current effect, since the spin Hall angles of Pt and W are opposite in sign [43]. The shapes of voltage curves for Pt/VO$_2$ and W/VO$_2$, accounting for the sign change, are very similar, indicating they are of the same origin.

At temperatures below 5 K, by further increasing $B$ above some certain field, $V_{2\omega}$ starts to decrease, showing a $B$-induced reduction of the paramagnetic SSE in the Pt-based device, which is not due to the magnetoresistance of Pt wire (Fig. 6c). A similar $B$-induced reduction was also observed in measurements on GGG [20,21]. This was interpreted [29] as the suppression of the interfacial spin-flip scattering between the Pt conduction electrons and the spin in the insulator, since at high fields and low temperatures, the Zeeman energy ($g\mu_B B$) of the spin becomes comparable to the thermal energy ($k_B T$).

The $2\omega$ signal has the orientation dependence of ***B*** in the film plane as expected for the spin Seebeck effect. As shown in Fig. 2d, at fixed field magnitudes |***B***| = 1 T and 6 T, the signal is described well by a $\cos\alpha$ dependence (the dashed curve), as expected for the SSE. The $2\omega$ voltage signal likewise depends linearly on the heater power at fixed ***B*** oriented at $\alpha = 0°$ (Fig. 2e), as expected for a SSE signal. A potential confounding effect in this experimental geometry, the ordinary Nernst response of Pt (W), expected to be linear with the applied magnetic field, cannot explain the observed magnetic field dependence of V$_{SSE}$ shown in Fig. 2a-c. Furthermore, with a 10-nm-thick insulating SiO$_x$ layer inserted between the Pt and the VO$_2$ film, the signal was reduced of 2 orders of magnitude (Fig. 16), consistent with the ordinary Nernst response measured in similar geometry [44].

We compare the magnitude of SSE in VO$_2$ with that in the ferrimagnetic insulator Y$_3$Fe$_5$O$_{12}$ (YIG). The spin Seebeck coefficient [45,46] is found as $\sigma_{SSE} = (V_{SSE}/l)/(dT/dz)$, where $l$ is the length of the ISH detector, and $dT/dz$ is the temperature gradient in the insulator. At low temperatures in bulk YIG, the measured σ$_{SSE}$ is around 5 µV/K [46]. For VO$_2$, a rough estimate of the thermal conductivity gives 10 nV/K at 8 T and 10 K, close to that estimated in YIG of 70 nV/K for 250 nm thickness and the same temperature range (see App. D, L). Given the uncertainties associated with interfacial thermal resistances, an alternative approach to comparing SSE responses between materials uses the spin Seebeck resistivity [39], defined as $R_{SSE} = (V_{SSE}/l)/j_Q$, where $j_Q$ is the heat flux through the insulator. In YIG, $R_{SSE}$ is ~ 10 nm/A with 100 nm thick at 10 K [39]; and in 100 nm-thick $VO_2$, $R_{SSE}$ is ~ 65 nm/A at 10 K and 8 T. That $VO_2$ has a SSE response comparable to that in the ferrimagnet YIG is striking, given that monoclinic $VO_2$ is not expected to host magnetic excitations.

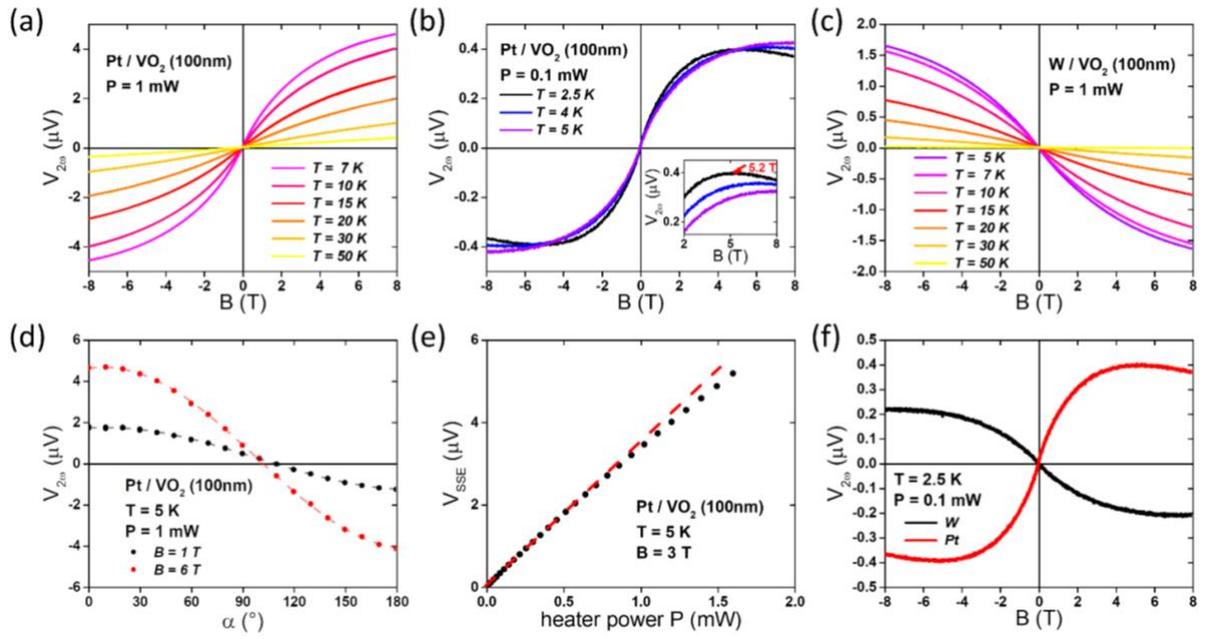

**Fig. 2**. (a-c) The second harmonic voltage as a function of field ($\alpha = 0°, \boldsymbol{B}||y$) at various temperatures for Pt (a,b) and W (c) detector wires on 100 nm thick $VO_2$. For Pt wire, data above 5 K are taken at 1 mW heater power; data at 5 K and below are taken at 0.1 mW heater power to minimize differences between local temperature and cryostat temperature. For W wire, all the data are taken at 1 mW. (d) Dependence of signal in Pt wire at 5 K with 1 mW heater power on in-plane field angle $\alpha$, showing expected cosine dependence. The device is misaligned in the plane by a few degrees relative to the ideal positioning. (e) Dependence of the spin Seebeck voltage on the heater power at 5 K and 0°. The SSE voltage is defined as the difference of the second harmonic signals between zero-field and 3 T. The slight sublinear dependence at high heater powers indicates a discrepancy between the local sample temperature and cryostat temperature. (f) Comparison between voltage responses of Pt/$VO_2$ and W/$VO_2$ devices at 2.5 K with an applied heater power of 0.1 mW.

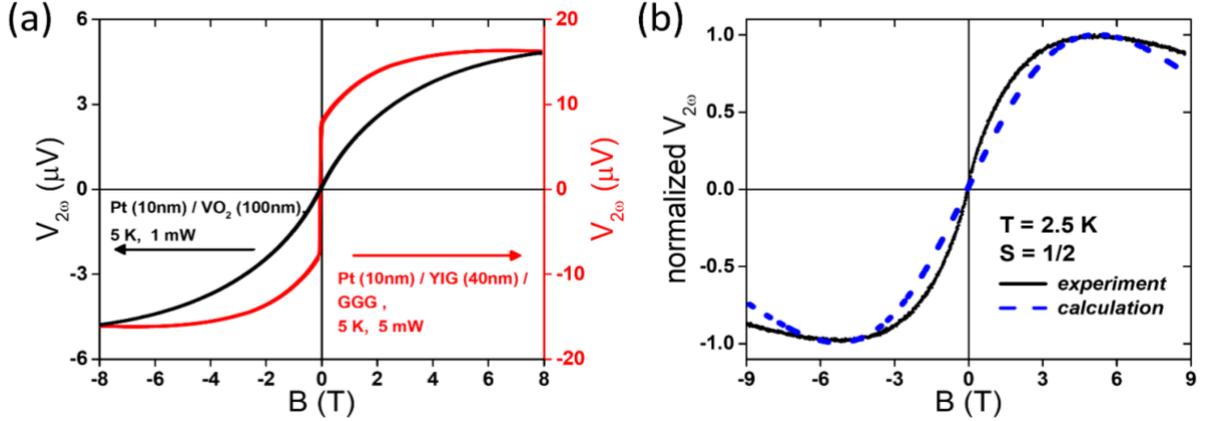

**Fig. 3**. (a) The second harmonic voltage as a function of field for Pt/VO$_2$ device and Pt/YIG/GGG device. (b) Comparison of normalized spin Seebeck voltage between experiment in Fig. 2b and the theoretical model calculation described in the text. The experiment data is taken at 2.5 K with 0.1 mW heater power. We obtained the optimal $\Theta_{CW}$ = -1.93 K by fitting at this temperature, consistent with a tendency toward antiferromagnetism. The data are normalized so that the maxima are set to 1.

The presence of a strong low temperature spin Seebeck response in VO$_2$ raises the question regarding the nature of the angular momentum-carrying excitations. The fact that the ground state of monoclinic (M1) VO$_2$ is a singlet-dimer state leads to considering whether thermally excited triplets ("triplons") may transport spin angular momentum, leading to a triplon SSE. A triplon SSE has previously been reported in the LSSE measurement configuration in the spin-Peierls system CuGeO$_3$ [27], where Cu atoms form one-dimensional spin-1/2 chains with antiferromagnetic exchange interactions. A key distinguishing feature of the triplon SSE is its voltage sign, consistent with the triplon current carrying magnetization in the same orientation as the bulk magnetization. In coerced ferromagnets (or paramagnons in paramagnets), conversely, a magnon transports magnetic moment that is antiparallel to the bulk magnetization. In the CuGeO$_3$ system [27], consistent with triplons as the spin-carrying excitations, the LSSE voltage is found to be of the opposite sign as the LSSE voltage in YIG, in which magnons provide the SSE response [10]. To test the LSSE voltage sign in our system, we made an analogous device on a YIG thin film of 40 nm thickness deposited on a GGG substrate. The sign of the LSSE signal in VO$_2$ devices is the *same* as that of the magnon-mediated SSE in YIG/GGG (Fig. 3a), in contrast to the CuGeO$_3$ case, seemingly ruling out the possibility that the SSE in VO$_2$ devices is caused by a current of triplons.

Assuming the ideal singlet dimer picture of the monoclinic VO$_2$ state, there should be no free magnetic moments. In the CuGeO$_3$ case, the free spin density (due to local disorder preventing singlet dimer formation) is estimated to be 0.02%, and the average distance between free spins is estimated to be around 1.5 μm, too dilute for correlations between the free spins to contribute to spin current transport [27]. In the VO$_2$ case, one analysis based on the low–$T$ susceptibility roughly estimates that ~15% V$^{4+}$ ions could be "free" ions residing in the otherwise dimerized system [37], though sample preparation would likely affect this greatly. For example, internal stresses in the film could potentially stabilize regions of two other insulating metastable phases of monoclinic, M2 (space group C2/m) and triclinic, T (space group P$\bar{1}$) by introducing tensile strain along the V-V zigzag chain [47], both of which could create some undimerized V ions. Deviations from ideal oxygen stoichiometry could likewise lead to unpaired spins. Further experiments involving radiation damage or other means of breaking V-V dimers could test this idea.

We consider whether the LSSE data from VO$_2$ can be understood within a particular model of the paramagnetic SSE due to the spin-flip scattering via the interfacial exchange coupling between localized moments in the VO$_2$ and conduction-electron spins in the Pt [29]. Within this model, the ISHE-induced voltage, $V_{SSE}$, can be expressed as

$$V_{SSE}/V_{SSE}^{max} = C \frac{SB_S(\xi)\xi^2}{\sinh(\xi/2)^2},$$

where $C$ is a normalization prefactor, $B_S(\xi)$ is the Brillouin function of spin $S$, and $\xi = g\mu_B B/k_B T$ is the dimensionless ratio of the Zeeman energy to the thermal energy. $B$ is within the Curie-Weiss molecular field model $B_{eff} = [T/(T - \Theta_{CW})]B$, where $\Theta_{CW}$ is a possible Curie-Weiss temperature of VO$_2$. In the formula, the only free parameter is $\Theta_{CW}$ and we get the optimal value of $\Theta_{CW}$ by fitting, as shown in Fig. 3b. Comparing with the measured spin Seebeck signal, the calculation shows the observed field-induced reduction above a similar field. However, the zero-field slope in the calculations is smaller than the observed signal; in addition, the high-field reduction predicted by the calculations is larger than the reduction observed in the measured data. Attempting to fit the data at higher temperatures requires a temperature-dependent $\Theta_{CW}$, implying that other temperature-dependent physics is important.

Even allowing $\Theta_{CW}$ to vary with temperature or considering spin-1 as well as spin-1/2 moments, it is not possible to simultaneously fit the low-field slope and the high-field reduction in SSE vs. B dependence. The quantitative disagreement between the experiment and the calculation suggests that the measured signal is not caused by a pure interfacial effect. In fact, the sign reversal of nonlocal SSE on $VO_2$ [38] on 100-nm-thick films implies that there is a bulk contribution to the SSE due to the local chemical potential of the spin-carrying excitations [48].

To constrain the mechanism driving the spin Seebeck response in $VO_2$, we examined its temperature dependence. Fig. 4a shows the temperature dependence of LSSE voltage response in a Pt wire on 100 nm thick $VO_2$ with different fields, from 2 K to 50 K. At constant heater power, the LSSE voltage at each field increases with decreasing temperature, reaching a maximum at a peak temperature $T_{peak}$, and decreases with further decreasing temperature. The peak temperature increases with increasing fields (Fig. 4b), qualitatively consistent with the linear field dependence of $T_{peak} \approx g\mu_B B/k_B - |\Theta_{CW}|$ from the model [29]. However, the model does not fully account for the temperature dependence originating from the Kapitza thermal boundary conductance.

Between 15 and 50 K, the LSSE voltage in $VO_2$ varies approximately as $T^{-2}$. In contrast, previous work on the paramagnetic SSE in GGG showed a steeper power-law decay of the LSSE voltage at constant heater power, proportional to $V_{2\omega} \propto T^{-3.384}$ [20]. The argument was to be roughly consistent with a Curie-like temperature dependent magnetization $M \propto 1/T$ combined with the temperature-dependent thermal conduction of the crystalline insulator and the Kapitza thermal boundary conductance between the metal and the insulator ($\kappa \sim T^3$ for both). The considerably weaker temperature dependence observed here in $VO_2$ is thus surprising. Although the magnetization of $VO_2$ at low temperatures was reported to be unusual [31,34], we have been unable to measure directly *M(T,H)* or the thermal conductivity of these thin films. This discrepancy in temperature dependence suggests a potentially strong temperature dependence of the interfacial spin exchange coupling at the $VO_2$/metal interface.

The spin-gapped system $Pb_2V_3O_7$ shows a similar peak behavior [28], attributed to the competition between the decreased paramagnon density and the increased paramagnon lifetime as the temperature decreases, the same explanation as argued in the ferromagnetic SSE [7]. In recent work in ferromagnets, however, both experiment [46] and theory [8], show that, at low temperatures the SSE can be dominated by a phonon-drag mechanism, where the spin current is induced by temperature-gradient-driven phonons via magnon-phonon interactions. In this case, the phonon-drag model predicts $V_{SSE} \propto \kappa \nabla T = j_Q$, which is constant in our measurement method, contrary to the observed temperature dependence.

As mentioned above, both bulk contributions [6] and the interfacial contributions [29] to the spin Seebeck response exist. We fabricated devices with the same geometry and fabrication protocol but varying thicknesses of $VO_2$ films [49]. The magnitude of the SSE response is expected to be directly proportional to the interfacial spin exchange coupling at the SOC metal/insulator interface, and thus extremely surface sensitive. The field dependence of LSSE voltage at 2 K shows no systematic trend of the magnitude with film thickness (Fig. 4c), while interfacial temperature differences should be governed by differences in the sound speed between the metal and the insulator and are not expected to vary by large amounts. This implies that the interfacial spin exchange can vary from device to device, even with nominally identical processing steps. When normalizing to its maximum value (Fig. 4d, and Fig. 10b), the normalized LSSE voltage as a function of field shows consistent behavior across all devices, implying an intrinsic mechanism in $VO_2$ related to its magnetization.

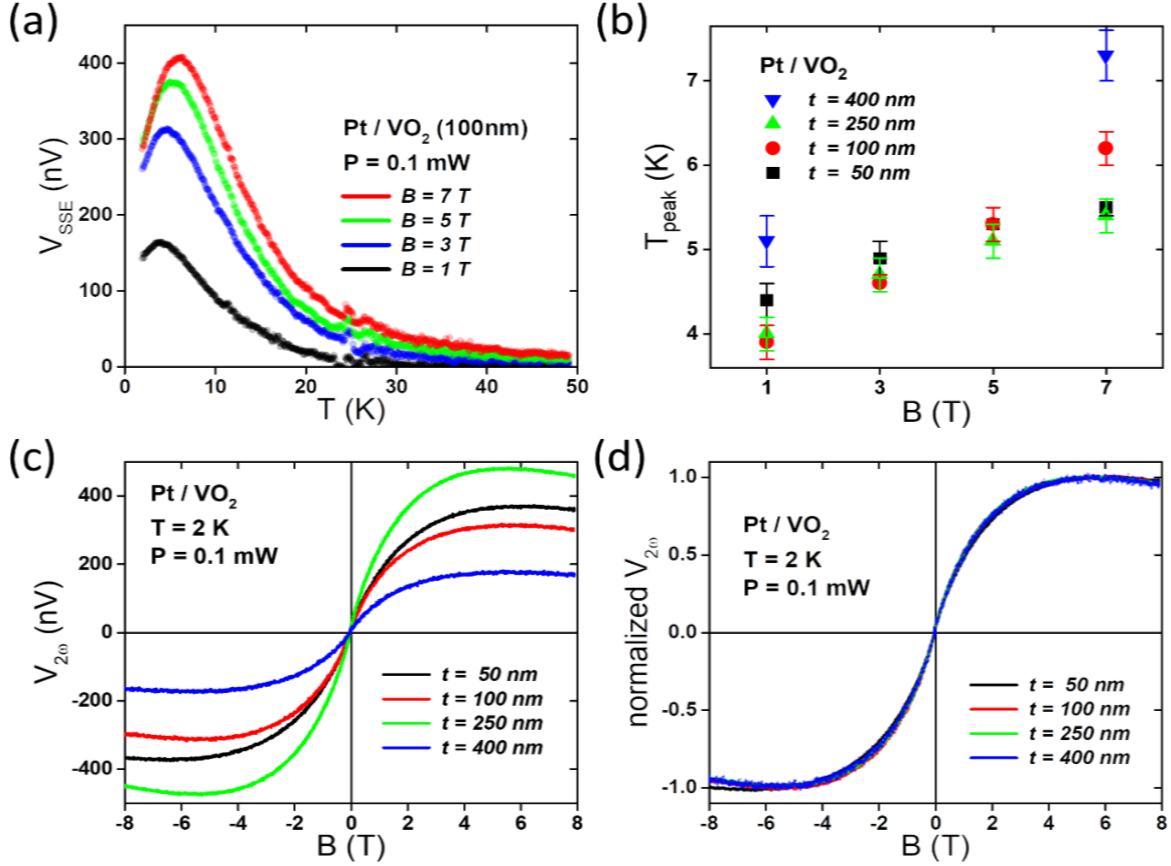

**Fig. 4.** (a) Temperature dependence of the LSSE voltage, defined as the difference between the second harmonic voltages at a certain field and 0 T, in the Pt wire on the 100-nm-thick $VO_2$ film at constant heater power of 0.1 mW and $\alpha = 0°$. (b) Field dependence of the peak temperature for different thicknesses of $VO_2$. For each thickness, the peak temperature increases with the field approximately linearly. (c) The field dependence of the second harmonic voltage for different thicknesses in the Pt/$VO_2$ device at 2 K for the different thicknesses shows no systematic trend of the magnitude of the LSSE voltage with film thickness. (d) When normalized to the maximum value for each film, the second harmonic voltage shows essentially identical dependence on the field, indicating a consistent mechanism associated with the $VO_2$ material.

## IV. CONCLUSIONS

We find a strong, temperature-dependent local spin Seebeck response in thin films of $VO_2$, comparable to that seen YIG, even though stoichiometric $VO_2$ is expected to be magnetically inert. The sign of the measured LSSE voltage is incompatible with thermally activated triplons as the spin-carrying excitations. While a recent model [29] of an interfacial SSE between a paramagnetic insulator and the strong spin-orbit metal is qualitatively consistent at fixed temperatures with the nonmonotonic field dependence observed at the lowest temperatures, the temperature and field dependence of the data and prior nonlocal measurements [38] support a bulk SSE interpretation. Additional studies of paramagnetism in the monoclinic phase of $VO_2$ are required to resolve the nature of spin transport in this correlated system.

**ACKNOWLEDGMETS** We thank E. Morosan for the use of the DynaCool instrument. We thank D. Cahill for the discussion of the thermal conductivity techniques at low temperatures. All work at Argonne, including fabrication of the YIG/GGG device was supported by the U.S. Department of Energy, Office of Science, Basic Energy Sciences, Materials Sciences and Engineering Division. The use of facilities at the Center for Nanoscale Materials, an Office of Science user facility, was supported by the U.S. Department of Energy, Basic Energy Sciences, under Contract no. DE-AC02-06CH11357. RL, TJL, and DN acknowledge support from DMR-2102028 for spin Seebeck measurement capabilities. DN and LC acknowledge support from DOE DE-FG02-06ER46337 for studies of $VO_2$ nanostructures. Synthesis, characterization, joint design of the experiments, extensive discussions, and joint writing of the manuscript (HN, ACB, IKS) were funded by the U.S. Department of Energy, Office of Science, Basic Energy Sciences under Award No. DE-FG02-87ER45332.

*Authors Declarations*

*Conflict of Interest*

The authors have no conflicts to disclose.

*Data Availability*

The data that support these findings are available on Zenodo (https://doi.org/10.5281/zenodo.10965577).

**APPENDIX A. MAGNETIZATION OF COMMERCIAL $VO_2$ POWDER**

Ideally, we need to measure the magnetization of $VO_2$ thin film in our device. However, due to the small thickness compared to the diamagnetic substrate sapphire (hundreds of nanometers compared to millimeter), the magnetic signal of $VO_2$ is overwhelmed by the diamagnetic background of the sapphire. For an example of $VO_2$ response at low temperatures, we measured a sample of commercially available VO$_2$ powder. In Fig. 5a, we show the field dependence of magnetization. No hysteresis is observed, implying that VO$_2$ is paramagnetic. The susceptibility increases when the temperature is lowered. The $1/\chi$ vs. $T$ plot shows the deviation from a straight line, indicating other paramagnetic contributions, rather than Curie's law, dominate at low temperatures. Extrapolating the high temperature trend implies a negative Curie-Weiss temperature, ~ -5.60 K.

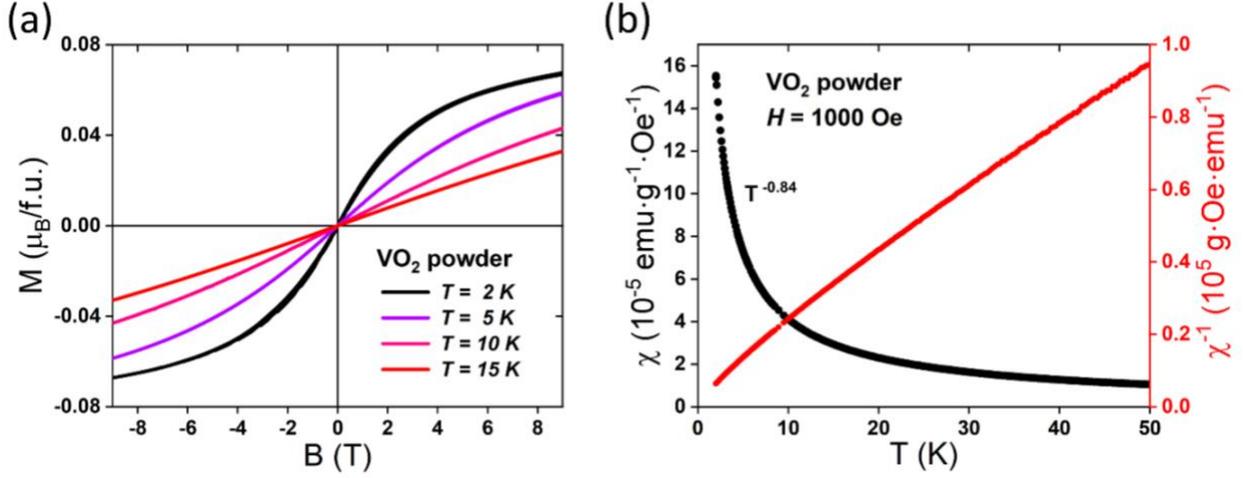

**Fig. 5.** (a) Field dependence of the magnetization of VO$_2$ powder, showing paramagnetism over the temperature range of interest. (b) Temperature dependence of magnetic susceptibility $\chi$ and $1/\chi$ in the field of 1000 Oe. The susceptibility shows non-Curie behavior at low temperatures.

**APPENDIX B. RESISTANCE OF THE PLATINIUM AND TUNGSTEN WIRE**

Since the spin Seebeck response is proportional to the resistivity of the spin-orbit metal, a change of the Pt and W resistance with temperature or field will affect the measured spin Seebeck voltage extrinsic to the actual spin Seebeck physics. Fig. 6a (6b) shows the temperature dependence of the resistance of the Pt (W) wire. The change of $R_{Pt}$ and $R_W$ in the temperature range from 50 to 5 K is relatively small, less than 2%. The field dependence of R$_{Pt}$ and R$_W$ at T = 5 K at some selected angles are shown in Figs. 6c and 6d, respectively. R$_{Pt}$ and R$_W$ change less than 0.1 % up to 8 T. In short, the contribution of the resistivity change in the Pt and W wires within the experiment's temperature and magnetic field ranges is negligibly small compared to the observed SSE signal in our devices.

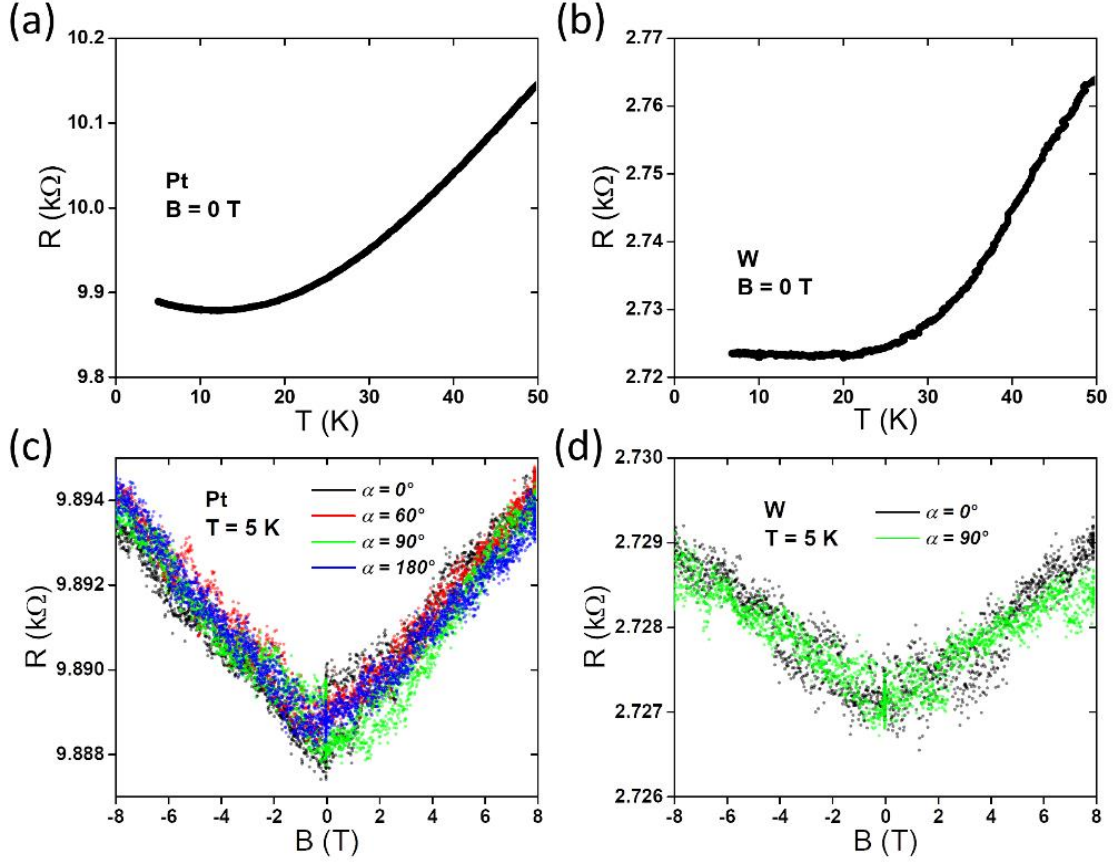

**Fig. 6.** Resistances of the Pt wire and W wire. (a, b) Temperature dependence of the resistances of the Pt and W wires. (c, d) The magnetoresistances of the Pt and W wires at 5 K, at some representative angles. The resistance changes for both metals with fields up to 8 T are below 0.1 % and is qualitatively consistent with weak antilocalization [50,51].

**APPENDIX C. ANGLE DEPENDENCE OF THE SSE**

Fig. 7a shows the field dependence of the second harmonic signal at T = 5 K with different in-plane field orientations. The sign of the signal is opposite for 0° and 180°, and the signal at 90° is almost zero, consistent with the expected symmetry of ISHE and the device geometry. Fig. 7b shows the temperature dependence of the SSE response at different angles. The temperature where the response reaches the maximum is independent of angle, and the amplitude of the signal scales as cos α, as expected. To show this more readily, we normalized the response to set the maximum to 1 and found that the SSE responses at different angles lie on the same curve (Fig. 7c). To conclude, the change of angle only affects the overall magnitude of the SSE response. These dependences are consistent with what is expected for the spin Seebeck effect.

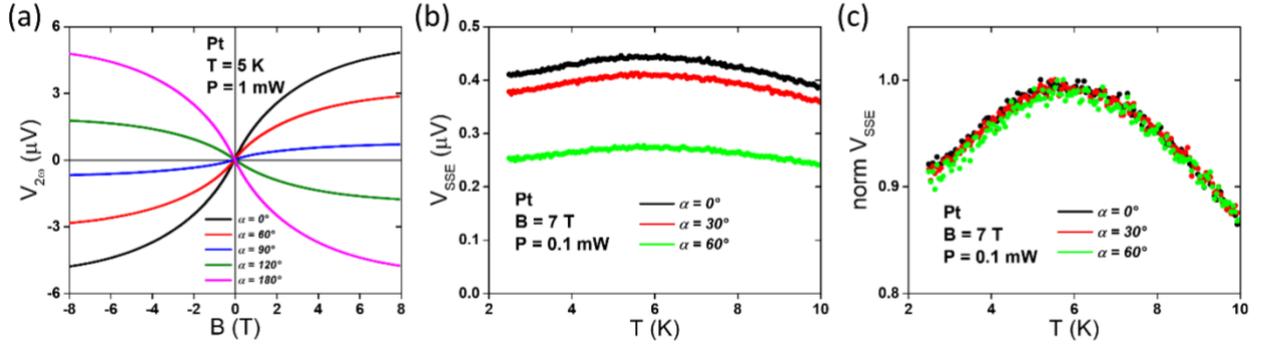

**Fig. 7.** (a) The second harmonic voltage as a function of field at 5 K for Pt at various in-plane field orientations. (b) Temperature dependence of the LSSE voltage, defined here as the difference between the second harmonic voltages at 7 T and 0 T, in the Pt wire at constant heater power of 0.1 mW and different angles. (c) The same data when the maximum value of the voltage in (b) is normalized to 1.

**APPENDIX D. ESTIMATION OF SPIN SEEBECK COEFFICIENT AND SPIN SEEBECK RESISTIVITY IN VO$_2$**

The comparison of spin Seebeck effect between different materials and the quantitative extraction of the precise thermodynamic coefficient ideally require knowledge of the exact temperature profile across the full stack, which, in our case, is Au-SiO$_x$-Pt-VO$_2$-sapphire-cryostat stack. This information, however, is extremely difficult to obtain in general, especially for thin film samples and across buried dielectric interfaces. There is no natural, reliable way to measure the temperature of the sapphire adjacent to the sample board, and similarly there is no way to measure any interfacial temperature difference at the boundary between the VO$_2$ films and the underlying sapphire, or between the Pt ISH detector and the VO$_2$ film.

There are two main approaches for quantitative comparisons of the magnitude of the LSSE response between different materials and experimental setups. One figure of merit is the actual spin Seebeck coefficient [45,46], $\sigma_{SSE} = (V_{SSE}/l)/(dT/dz)$, where $l$ is the length of the Pt detector, and $dT/dz$ is the temperature gradient through the SSE material along the direction of heat flow. An alternative figure of merit, formulated knowing that interfacial temperature differences can be relevant and are difficult to measure, is the spin Seebeck resistivity [39], defined as $R_{SSE} = (V_{SSE}/l)/j_Q$, where $j_Q$ is the heat flux through the SSE insulator. Below we estimate both $\sigma_{SSE}$ and $R_{SSE}$ for the Pt/VO$_2$ devices and find responses comparable to what is observed in ordered magnetic material such as YIG.

We roughly estimate the temperature gradient across the VO₂ film in a typical device, given by $dT/dz = \dot{q}/(\kappa_{VO2} A)$, where $\dot{q}$ is the heater power transported vertically through the Pt/VO₂ interface, $\kappa_{VO2}$ is the VO₂ thermal conductivity, and $A$ is the cross-sectional interface area for the transport. Finite-element thermal modeling (see App. K) supports the conjecture that in our measurement setup, for reasonable values of thermal boundary resistance parameters, the heat flux through the VO₂ film is approximately constant as a function of temperature, which means the dominant thermal path for power generated in the heater is downward through the VO₂ film. Note that the Pt detector wire is 800 μm long while the Au heater wire is 1300 μm long; thus, for a total heater power of 1 mW in Au wire, at most about 0.615 mW of the heater power is transported downward through the Pt wire; larger thermal boundary resistances would reduce this fraction.

To the best of our knowledge no data is available for the cross-plane low temperature thermal conductivity $\kappa_{VO2}$ of VO₂. Directly measuring the low temperature cross-plane thermal conductivity of the VO₂ films is very difficult. The most common approach (the 3ω method [52]) is not applicable at low temperatures because the T-dependence of the typical heater material (Pt) resistivity vanishes below about 20 K (that is, $dR/dT \rightarrow 0$). Optical techniques [53] that rely on thermal expansion of the film material similarly do not perform at low temperatures because the temperature-dependent thermal expansion coefficient is suppressed at low temperatures. We can get a rough estimate of the thermal conductivity from the kinetic theory approach, using $\kappa_{VO2} = (1/3) C v_s l_m$, where $C$ is the temperature-dependent specific heat per unit volume, $v_s$ is the transverse speed of sound, and $l_m$ is an effective phonon mean free path. This assumes phonon diffusion, so self-consistency would require it to be applied to films thicker than phonon mean free path and thicker than a typical thermal phonon wavelength. At 10 K, specific heat of VO₂ was reported to be 15.4 mJ/(mol·K) [54], and converting into per-unit-volume, 848 J/(m³K). A reasonable speed of sound is 4500 m/s [55], giving a thermal phonon wavelength at 10 K of about $h v_s / k_B T = 22$ nm. For consistency with the idea of diffusive phonon conduction, we can assume a phonon mean free path smaller than the film thickness; should the phonon mean free path be comparable to the film thickness, the cross-plane thermal conductivity would be larger by up to a factor of order 3. Assuming a thermal phonon mean free path of 100 nm and diffusive phonon transport implies a $\kappa_{VO2}$ thermal conductivity close to 0.13 W/m·K at 10 K. Then given $A = 8 \times 10^{-9}$ m², this would imply a temperature gradient across a 250-nm-thick film of $\frac{dT}{dz} = 5.91 \times 10^4$ K/m at a sample temperature of 10 K with applied total heater power of 0.1 mW. (Thermal boundary resistances would reduce this thermal gradient by favoring lateral heat conduction out of the heater, rather than vertical heat transport. Thus, the estimates of $\sigma_{SSE}$ and $R_{SSE}$ that we find here are likely underestimates.)

Using the temperature gradient in VO$_2$, we can then compare SSE in VO$_2$ and YIG in terms of spin Seebeck coefficient $\sigma_{SSE} = (V_{SSE}/l)/(dT/dz)$, where $l$ is the length of the Pt detector, and $dT/dz$ is the temperature gradient estimated above. In 250-nm-thick VO$_2$, $V_{SSE}$ is ~ 500 nV at 10 K and 8 T, and giving and estimated $\sigma_{SSE}$ of 10 nV/K. At low temperatures in YIG, $\sigma_{SSE}$ is measured around 5 μV/K for bulk [46]. Considering the thickness dependence of the magnon SSE [7, 8], and the magnon diffusion length in 210-nm-thin YIG at 10 K was reported to be 8 μm [56], the coefficient $\sigma_{SSE}$ is estimated to be 70 nV/K in YIG for a film 250 nm thick. This differs from the VO$_2$ estimate only by a factor of 7; a larger VO$_2$ thermal conductivity and important thermal boundary resistances would imply a larger estimated $\sigma_{SSE}$ for VO$_2$, closer to the YIG value.

Given the uncertainties associated with interfacial thermal resistances and the difficulty in measuring temperatures of every material at each interface, an alternative approach to comparing SSE responses between materials uses the spin Seebeck resistivity [39], $R_{SSE} = (V_{SSE}/l)/j_Q$, where $j_Q$ is the heat flux through the insulator. In YIG, $R_{SSE}$ was reported to be ~ 10 nm/A with 100 nm thick at 10 K [41]; and in 100-nm-thick VO$_2$, using the heat flux computed from a total heater power of 0.1 mW and the device dimensions, $R_{SSE}$ is ~ 65 nm/A at 10 K and 8 T, even larger than that in YIG.

In summary, the magnitude of the local SSE response in VO$_2$, a nominally non-magnetic material, is comparable in magnitude to the SSE response of YIG thin films.

**APPENDIX E. TEMPERATURE DEPENDENCE OF THE LSSE**

Fig. 8a shows the temperature dependence of LSSE response in another device with Pt wire on a 100-nm-thick VO$_2$ film at 1 and 7 T, from 2.5 K to 50 K. This device doesn't show peak behavior at 1 T, different from the device in the main text (Fig. 4a). We attribute this to variation in the Pt/VO$_2$ interfacial quality. Fig. 8b shows the temperature dependence of LSSE response in W wire at 1 and 7 T, from 2.5 to 50 K. When the heater power is held constant, the magnitude of LSSE voltage at 1 T increases with decreasing temperature, whereas the LSSE voltage at 7 T reaches the maximum at ~ 8.1 K. The shape is qualitatively similar to that in the Pt wire.

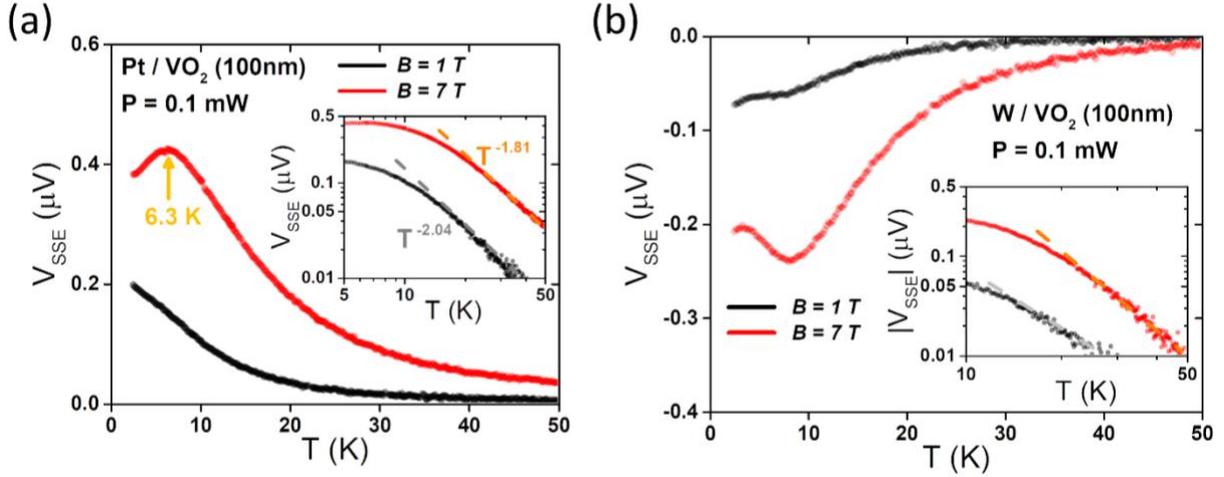

**Fig. 8.** Temperature dependence of the LSSE voltage, defined as the difference between the second harmonic voltages at the indicated field and 0 T, in the Pt wire (a) and W wire (b) on a 100-nm-thick $VO_2$ film at constant heater power of 0.1 mW and $\alpha = 0°$.

## APPENDIX F. ADDITIONAL DATA ABOUT THE SSE FOR DIFFERENT THICKNESSES OF $VO_2$ FILM

Fig. 9 shows the field and temperature dependence of the second harmonic signal at different temperatures for different thicknesses of $VO_2$ film. The responses are qualitatively all very similar; quantitative comparisons are shown in Fig. 4 of the main text and in Fig. 10 below. Fig. 10 shows the field dependence at 5 K for different thicknesses of $VO_2$ film. Similar to that at 2 K in Fig. 4c,d in the main text, the magnitude doesn't show a systematic trend with thickness, and the field dependence is quite identical.

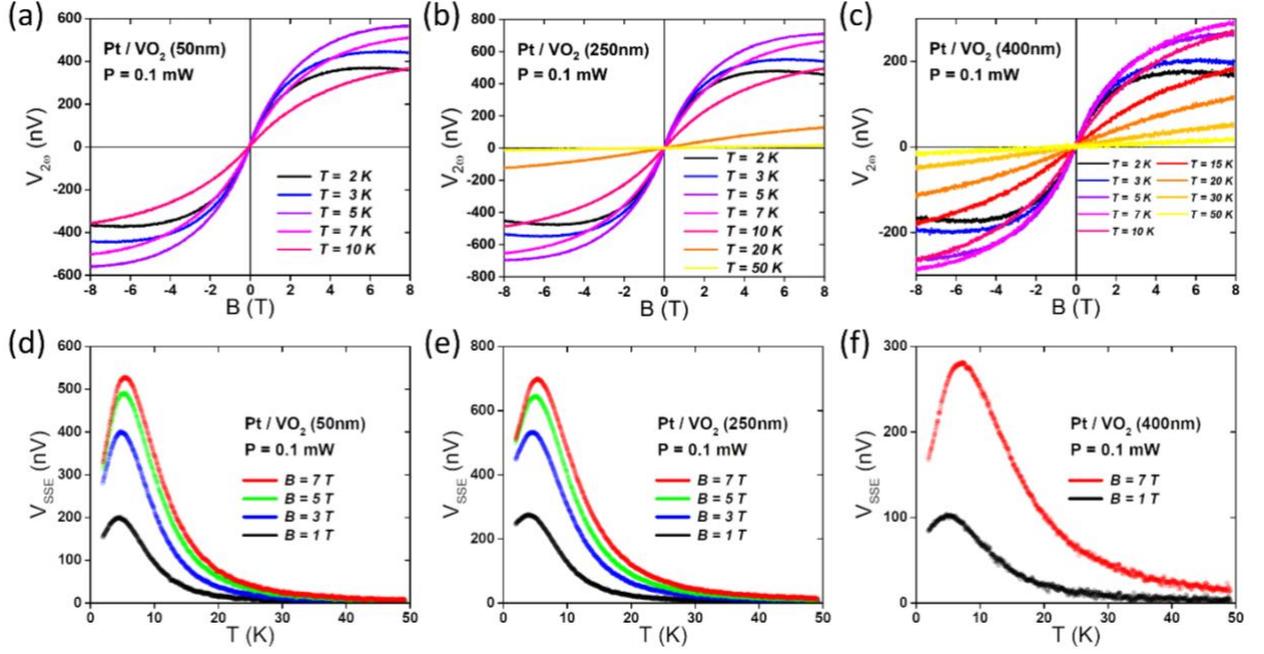

**Fig. 9.** (a-c) The field dependence of the second harmonic voltage for other film thicknesses (50 nm, 250 nm, 400 nm) in Pt/VO$_2$ devices at different temperatures. (d-f) The temperature dependence of the LSSE voltage for other film thicknesses (50 nm, 250 nm, 400 nm) in Pt/VO$_2$ devices at different fields.

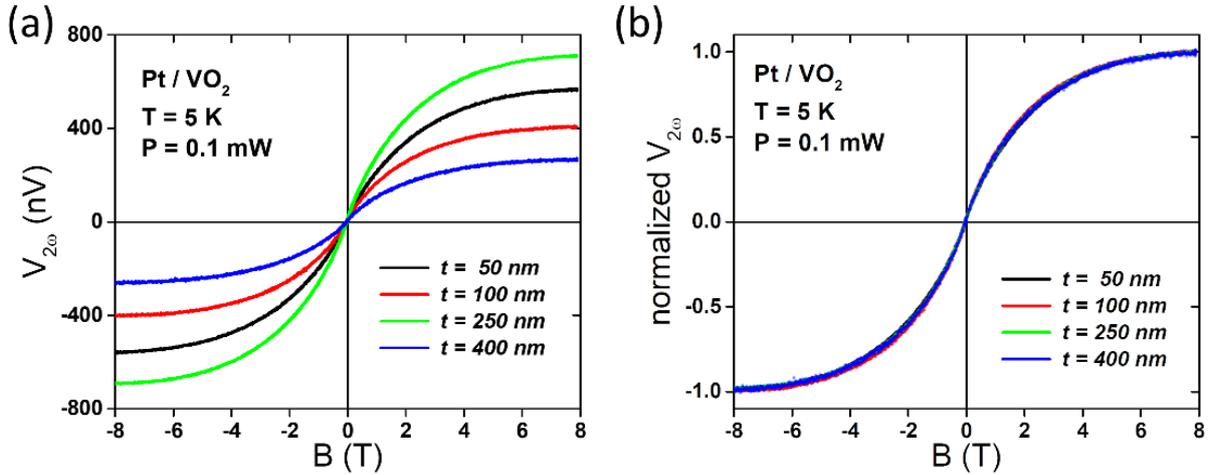

**Fig. 10.** (a) The field dependence of the second harmonic voltage for different thicknesses in Pt/VO$_2$ device at 5 K. (b) The same dataset in (a) with normalization.

## APPENDIX G. EFFECT OF THE HEATER POWER

The driving force of SSE, either the temperature gradient across the bulk of the VO$_2$, or the temperature difference at the interface between VO$_2$ and Pt, is proportional to the heater power. In the absence of self-heating effects, it is expected that the signal should fall on the same curve when normalized to the heater power. However, our observations indicate that self-heating can play a role at high heater powers. Fig. 11 shows the field dependence of the second harmonic signal at 1.8 K with different heater powers in the 14T-DynaCool. With increasing the power, the field where the signal reaches the maximum gets larger. A simple explanation is that the high heater power inevitably increases the temperature of the Pt and $VO_2$ significantly above the cryostat temperature, and then a larger field is needed to let the Zeeman energy balance the thermal energy.

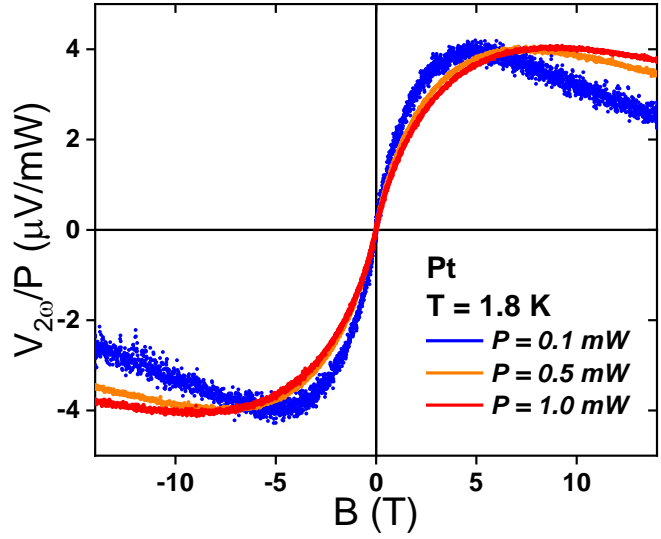

**Fig. 11.** Field dependence of the ratio of second harmonic voltage to the heater power at a cryostat temperature of 1.8 K for three heater powers. The trend here is consistent with the higher heater powers elevating the local Pt temperature significantly above the cryostat temperature.

**APPENDIX H. DETAILS OF THE FIELD-DEPENDENCE FITTING PROCEDURE**

As shown in the main text, the ISHE-induced voltage, $V_{SSE}$, can be expressed as:

$$v_{cal}(B) = V_{SSE}(B)/V_{SSE}^{max} = C \frac{SB_S(\xi)\xi^2}{\sinh(\xi/2)^2}, \qquad (1)$$

where C is a normalization prefactor, $B_S(\xi)$ is the Brillouin function of spin S, and $\xi = g\mu_B B/k_B(T - \Theta_{CW})$ is the dimensionless ratio of the Zeeman energy to the thermal energy, in which g is Landé g-factor, $\mu_B$ is Bohr magneton, $k_B$ is Boltzmann constant, $\Theta_{CW}$ is a possible

Curie-Weiss temperature of VO2. In the formula, the only free parameter is $\Theta_{CW}$, and to get the optimal value of $\Theta_{CW}$, we use the least squares fitting. We build the loss function as follows:

$$R^2 = \sum_i [v_{obs}(B_i) - v_{cal}(B_i)]^2 , \quad (2)$$

where $v_{obs}(B_i)$ is the observed normalized SSE response at field $B_i$, $v_{cal}$ is given by equation (1), and then find the value of $\Theta_{CW}$ to minimize $R^2$.

**APPENDIX I. MEASUREMENTS OF THE TEMPERATURE RISE OF THE FULL STACK**

We use the Johnson-Nyquist (JN) noise in the Pt detector itself under different heater powers to estimate quantitatively the temperature difference between the Pt and the cryostat. The details of the method have been reported elsewhere [44]. Fig. 12a shows the temperature rise $\Delta T_{Pt}$ (above the cryostat temperature measured using a Cernox thermometer) determined from JN noise in the Pt wire as a function of heater power at the cryostat temperature of 5 K for the 100-nm-thick VO2 film sample, while Fig. 12b shows the temperature dependence of $\Delta T_{Pt}$ at fixed heater power of 1 mW. $\Delta T_{Pt}$ grows linearly in the high heater power region and decreases with increasing temperature, similar to that observed in the Pt/SiO2 interface [44].

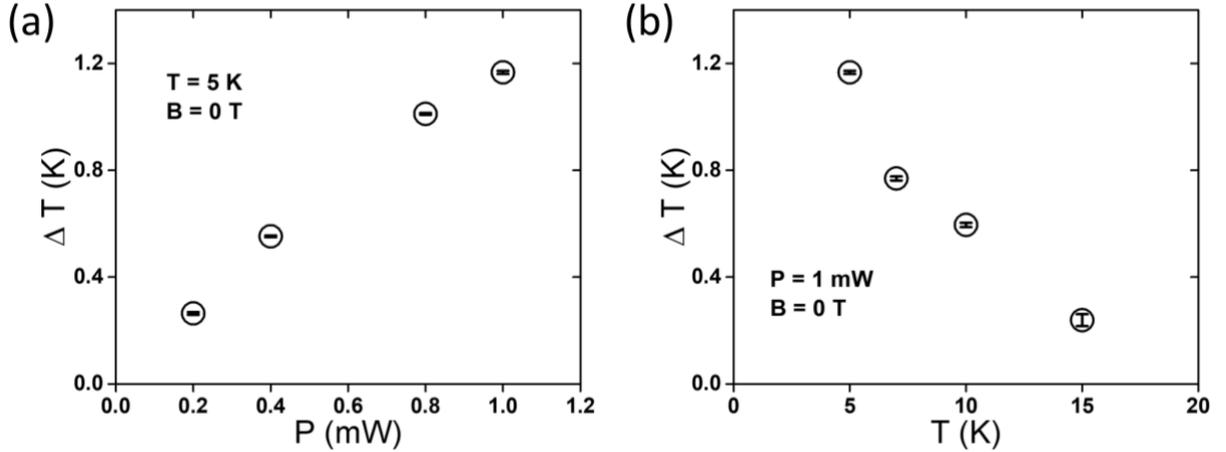

**Fig. 12.** (a) Temperature rise of Pt wire as a function of heater power at the cryostat temperature of 5 K as found via Johnson-Nyquist noise thermometry. (b) Temperature rise of Pt wire at fixed heater power of 1 mW as a function of cryostat temperature.

**APPENDIX J. OPTICAL IMAGE OF THE DEVICE**

Fig. 13 shows an optical image of the Pt/VO2 (100 nm) device in the main text.

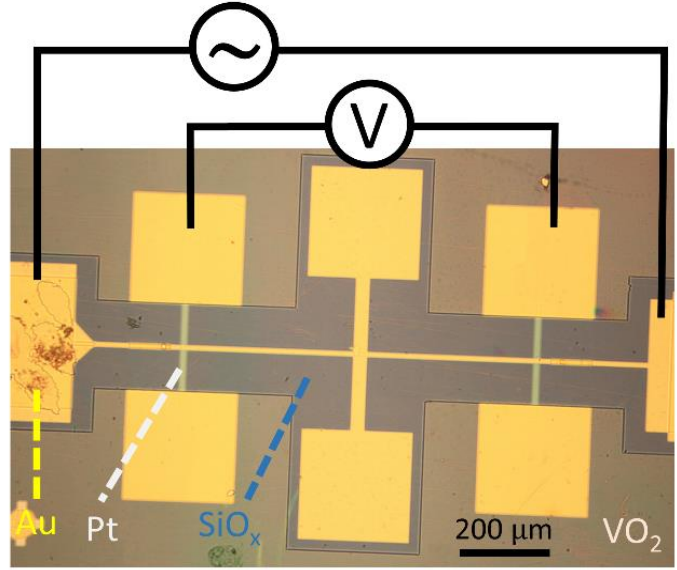

**Fig. 13.** Optical microscope image of a representative device.

**APPENDIX K. EXAMPLE DATA ON Pt/YIG/GGG AND Pt/GGG**

Fig. 3a of the main text compares the SSE response in a Pt/VO$_2$ device with that observed in a Pt/yttrium iron garnet (YIG) film/gadolinium gallium garnet (GGG) substrate device. The key takeaway from that comparison is that the response in VO$_2$ has the same sign (for the same wiring configuration) as the response in YIG, when the SSE in the latter is known to result from the propagation of magnons.

The SSE response in Pt/YIG/GGG device shown in Fig. 3a shows a sharp jump near zero field (the coercive field of the YIG film) and a more gradual additional response that resembles the expected magnetization vs. field for GGG. We note that this magnon-mediated LSSE signal shows contributions from the magnetization of both the YIG (the sharp jump near zero field) and the GGG (the high field variation), indicating that in that structure the LSSE is driven by the temperature gradient across the whole YIG/GGG stack, since for YIG of this thickness, the high-field suppression was not observed [59,60]. Fig. 14 shows representative example data taken in devices fabricated with a Pt electrode on YIG/GGG at high temperature (when the paramagnetic response of the GGG with field is comparatively weak) and with a Pt electrode on GGG at low temperatures, showing paramagnetic response very similar to that reported in Ref. [20]. The total response in the YIG/GGG stack structures involves both the direct YIG response and an additional contribution to the spin current in the YIG due to the GGG response. It is not immediately clear whether the GGG contribution originates more from the temperature gradient across the bulk GGG or more from an interfacial magnon temperature difference between the YIG and the GGG, and this is a challenging issue to resolve. Regardless, the ISH response of the Pt in these devices comes from processes involving magnons in the YIG, the material in direct contact with the Pt. The conclusion that the Pt/VO$_2$ SSE response has the same sign as that seen in a system governed by magnons is robust, seemingly ruling out mobile triplet excitations [27] as the origin of the SSE signal in VO$_2$.

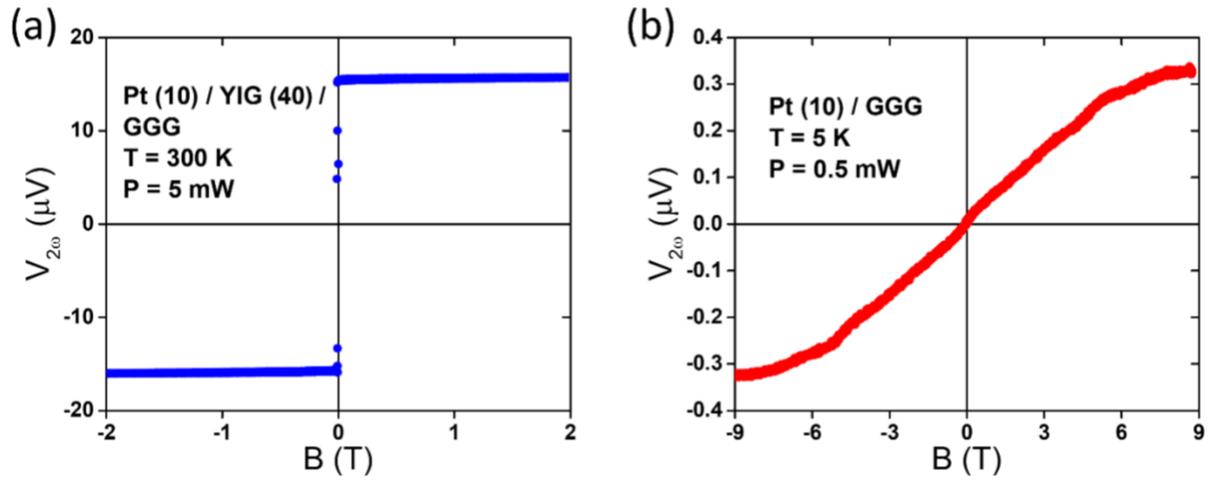

**Fig. 14.** (a) LSSE response of a Pt/YIG/GGG device measured at 300 K with 5 mW of heater power, showing the clear coercive switching of the YIG layer. The paramagnetic response of the GGG is small on this scale at this high temperature. (b) LSSE response of a Pt/GGG device measured at 5 K with 0.5 mW heater power, showing the paramagnetic SSE response of the GGG as in Ref. [20].

**APPENDIX L. THERMAL MODEL OF SSE DEVICE ON VO$_2$**

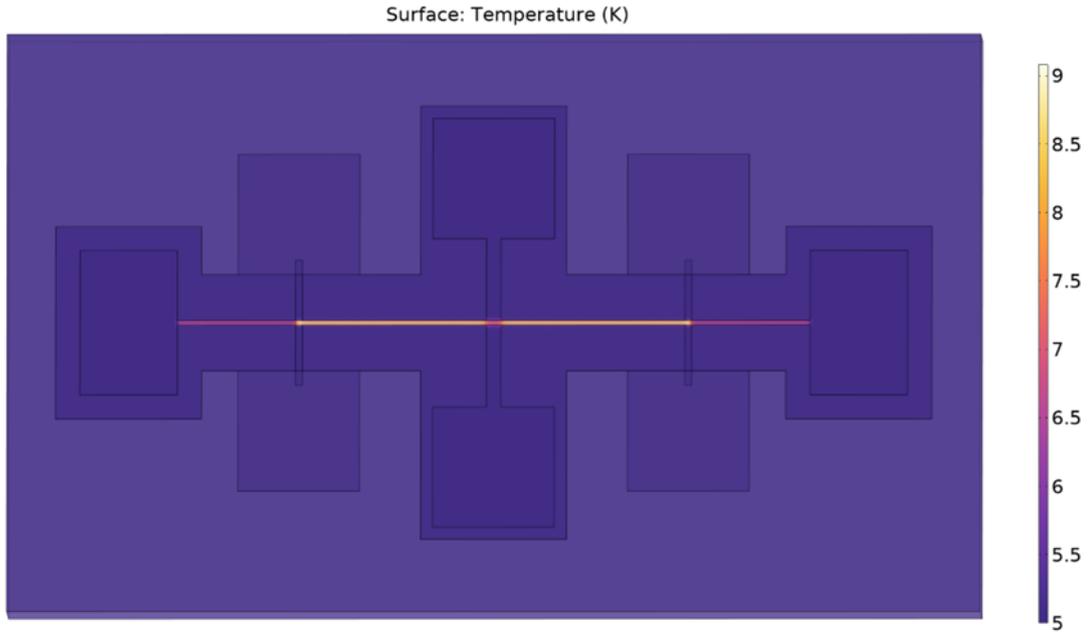

**Fig. 15.** Top view of the temperature profile of the thermal model for the SSE device on VO$_2$. Heat enters the Au heater wire and exits at the bottom of the substrate which is held to a "cryostat temperature" of 10 K. The Au wire is colder in the very center because of lateral conduction to nonheated Au contacts. The Au wire is also colder at its ends because there it is not over the Pt wire and thus the thermal path of the heat involves fewer thermal boundary resistances (the main thermal resistance contributors).

To aid in a quantitative estimate of the LSSE response (App. D), we constructed a finite element thermal model using COMSOL, in part to test the common assumption that all heat generated by the Au heater wire travels straight downwards, heating the area of the substrate directly below the Pt detector wire. To produce the thermal model, we used thermal conductivity values at 10 K for the VO$_2$ (estimated above in App. D), Al$_2$O$_3$ [57], and SiO$_x$ [58] of 0.13, 85, and 0.1 W/(m·K), respectively.

In addition to contributions from the bulk layers as above, the total thermal path between the Pt ISH detector and the cryostat also involves several thermal boundary resistances (BRs). The data in Fig. 12 demonstrate that these thermal resistances are not negligible, as to reproduce the directly measured temperature of Pt wire, the total thermal resistance from Pt to the substrate must be roughly two orders of magnitude larger than from the thermal resistance of the material layers themselves. BRs across metal/dielectric interfaces are expected to be the largest since the thermal conduction mechanism changes at those interfaces from electron-dominated to phonon transport, as well as due to acoustic mismatch between materials. Directly measuring these thermal BRs is very difficult in practice; worse still, they depend strongly on materials, interface quality, deposition method (e.g. evaporation vs. sputtering), process conditions, etc. Our approach is to insert thermal BRs at the metal/dielectric interfaces in the model and then vary them until the temperature of the Pt wire in the simulation equals the temperature of the Pt wire directly measured using noise thermometry, as stated in App. I. For a heater power of 1 mW, at cryostat temperatures of 5 K, 10 K, and 15 K, the Pt temperature is 6.2 K, 10.6 K and 15.3 K (Fig. 12). To reproduce these Pt wire temperatures in the simulation requires total thermal BRs of 18.9, 9, and 2.75 μK·m$^2$/W, respectively.

Quantifying the LSSE as either the SSE coefficient, $\sigma_{SSE}$, or the spin Seebeck resistivity, $R_{SSE}$, requires an estimate of the fraction of heat generated that is transported vertically through the insulating material below the Pt ISH detector. Assuming these thermal BRs make the model accurately describe the real device, we can then estimate how much heat travels down to the Pt wire and how much travels sideways in the SiO$_x$ layer. For cryostat temperatures of 5 K, 10 K, and 15 K, 78.3 %, 75 %, and 81.9 % of the heat current enters the top of the Pt wire (over 98 % of this heat current then enters the substrate directly below the Pt detector wire), and the rest travels sideways, not contributing to the detected spin Seebeck effect. If no thermal BRs are included, then 92 % of the area-estimated heat power (0.651 mW for 1 mW total heater power, as set by the Pt and Au geometry) reaches the top of the Pt wire and the Pt temperature only increases by 0.15 K regardless of cryostat temperature. This demonstrates that the heating of the substrate below the Pt detector wire is overestimated by assuming all generated heat travels directly downwards, and thus the VO$_2$ spin Seebeck coefficient and spin Seebeck resistivity estimated in the main text and S5 are underestimates.

**APPENDIX M. NERNST MEASUREMENTS ON A Pt/SiO$_x$/VO$_2$ CONTROL DEVICE**
To show the normal Nernst response is much smaller compared to the SSE response, and to confirm that the measurements reflect interfacial processes between the Pt and VO$_2$, we also characterized devices made with a 10-nm-thick insulating SiO$_x$ spacer layer inserted between the Pt detector wire and the 400-nm-thick VO$_2$ thin film, to block spin current. As shown in Fig. 16a, the signal is roughly linear to the applied field, similar to that was reported in Ref. [44]. This signal results from the Nernst-Ettingshausen response of the sputtered Pt material. The normalized signal in Fig. 16b shows that the SSE response is much larger than the Nernst response.

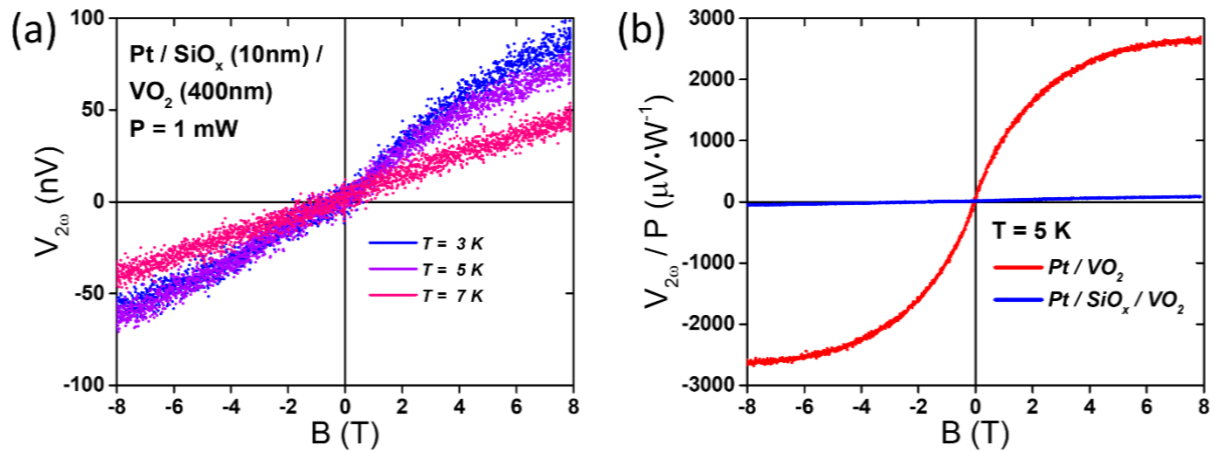

**Fig. 16.** (a) The field dependence of the second harmonic voltage in a Pt/SiO$_x$/VO$_2$ device at several temperatures. (b) Comparison of the normalized second harmonic voltage to heater power for the Pt/VO$_2$ device and a Pt/SiO$_x$/VO$_2$ device on the same VO$_2$ film at 5 K.